\documentclass[twocolumn,showpacs]{revtex4-1}
\usepackage{graphics,epsfig,graphicx}
\usepackage{color}

\usepackage{latexsym}
\usepackage{calrsfs}
\usepackage{changes}

\begin{document}



\title{Observation of macroscopic coherence in self-organized dipolar
excitons}

\author{M. Alloing$^{1}$, D. Fuster$^{2}$, Y. Gonz\' {a}lez$^{2}$, L.
Gonz\' {a}lez$^{2}$ and F. Dubin$^{1}$}

\affiliation{$^{1}$ ICFO-The Institut of Photonic Sciences,
Av. Carl Friedrich Gauss, num. 3, 08860
Castelldefels (Barcelona), Spain}
\affiliation{
$^{2}$ IMM-Instituto de Microelectr\'{o}nica de Madrid (CNM-CSIC), Isaac
Newton 8, PTM, E-28760 Tres Cantos, Madrid, Spain }

\date{\today }
\pacs{78.67.De, 73.63.Hs, 73.21.Fg, 78.47.jd}

\begin{abstract}

We report experiments showing that spatially indirect excitons confined
in a wide single quantum
well can exhibit macroscopic spatial coherence. Extended coherence is
spontaneously established in the regime where indirect excitons form a
distinctive
ring shaped pattern fragmented into microscopic beads. These contain 
a large concentration of indirect excitons at sub-Kelvin temperatures, the
excitons spatial coherence being the greatest in
the vicinity of the fragments.

\end{abstract}

\maketitle


Collective quantum states realized with  
semiconductor excitons have attracted a very large attention since 
seminal theoretical predictions were made in the 1960's
\cite{Blatt_1962,Keldysh_BCS,Keldysh_BEC}. To demonstrate such
phenomena
numerous systems were probed, from bulk materials to low
dimensional
heterostructures \cite{Snoke_Book}, and promising realizations of excitons
were then identified. These include notably para-excitons in
cuprous oxide \cite{Lin_1993,Snoke_1990,Kosuke_2011} but also spatially
indirect excitons confined in bilayer heterostructures
\cite{Timofeev_2008,High_2012,High_2012b,Chen_2006,Stern_2012}. 

Spatially indirect excitons are created by imposing a spatial
separation between the electrons and holes constituting excitons. This is
achieved for instance by applying an electric field perpendicular to a
single or a double quantum well. Thus, electrons and holes form two distant
layers and have wave-functions which experience a small overlap. Indirect
excitons thereby benefit 
from a lifetime that is long compared to the rapid thermalization to the lattice
(bath) temperature in this two-dimensional geometry \cite{Ivanov_2004}. In
addition, they exhibit a  large and well-oriented electric dipole
such that repulsive dipole interactions between excitons prevent the collapse
into a
plasma at large densities \cite{Butov_review}.

The unique physical properties of indirect excitons have allowed latest
experiments to cool dense ensembles ($\sim$10$^{10}$ cm$^{-2}$) to sub-Kelvin
bath temperatures \cite{High_2012b,Schinner_2011,Alloing_2011}. In
this regime, theoretical studies indicate that the excitons quantum statistics
shall be revealed \cite{Ivanov_2004} and for instance lead to the appearance of
macroscopic spatial
coherence \cite{Leggett_2006}. Remarkably, 
such a spontaneous buildup of macroscopic coherence has been reported
very recently by Butov and co-workers \cite{High_2012,High_2012b,Yang_2006} who optically
injected  indirect excitons in a double quantum well heterostructure. 

In this letter, we show that macroscopically coherent indirect
excitons can also form in a wide single quantum well. Precisely,
we report experiments where electron-hole pairs were optically injected and
subsequently diffused to form indirect excitons spontaneously arranged along
a ring-shaped pattern. Lowering the semiconductor bath temperature below a
few Kelvin, we observed that the exciton ring contracted spatially and also
underwent a fragmentation into microscopic
beads. The emission spectrum indicates that the beads contain a large
concentration of excitons at sub-Kelvin bath
temperatures ($\sim$10$^{10}$ cm$^{-2}$). Furthermore, interferometric
measurements reveal that the fragmentation of the exciton ring
coincides with a rapid increase of the emission's
spatial coherence. In the vicinity of the fragments, coherence extends
over a micrometer at 350 mK.

In the following we report studies of a 1 $\mu$m thick
field-effect device
where a 250 $\mathrm{\AA}$ wide
GaAs
quantum well is embedded. The quantum well is surrounded
by Al$_{0.3}$Ga$_{0.7}$As barriers and placed 900 nm below a uniform
semi-transparent gate
electrode deposited on the surface of the sample. For our measurements this
electrode was biased at a constant voltage V$_\mathrm{g}$=-4.17 V with respect
to the conductive sample's substrate that was grounded. Hence, minimum
energy states for electrons and holes are displaced in opposite directions and
the ground excitonic
transition is then spatially indirect. We optically injected 
electronic carriers in this device using a 
500 ns long laser excitation at 640 nm, i.e. at an energy slightly below the
Al$_{0.3}$Ga$_{0.7}$As bandgap \cite{Bosio_1988}. Then we studied 
the photoluminescence reemitted by the electron-hole bilayer at low bath
temperatures (T$_b\leq$ 7 K).
Details of our experimental apparatus may be found in Ref. \cite{Alloing_2012}.

\begin{figure}
\includegraphics[width=8.5cm]{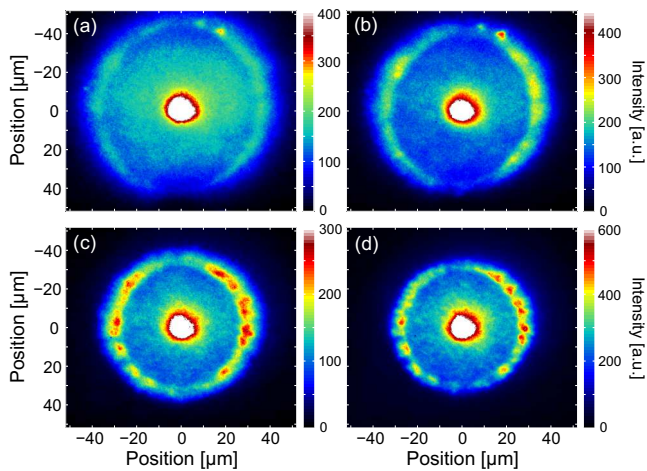}
\caption{(Color Online): Real space photoluminescence emitted by the
single
quantum well at low bath temperatures: from (a) to (d) T$_\mathrm{b}$ is set to
7, 5, 2.5 and 0.34 K.
The center of the emission pattern coincides with the region that is laser
excited, the laser spot extending over $\approx$ 10 $\mu$m
in these experiments. All measurements were acquired in a 40 ns time interval 20
ns after termination
of the laser excitation.}
\end{figure}

In Figure 1, we present the photoluminescence emission for
T$_\mathrm{b}\leq$ 7K.  In this temperature range, above a
threshold laser excitation we observed that a
ring-shaped photoluminescence formed at about 30 $\mu$m from
the illuminated region ($\approx$ 10 $\mu$m waist) where electronic
carriers were injected. As we lowered the bath temperature, we noted that the
ring pattern contracted and was marked further (Fig.2.b). Most notably, it also
underwent a spatial fragmentation into microscopic beads for $T_\mathrm{b}$ less
than $\approx$ 3K (see Fig. 1).

\begin{figure}
\includegraphics[width=8.5cm]{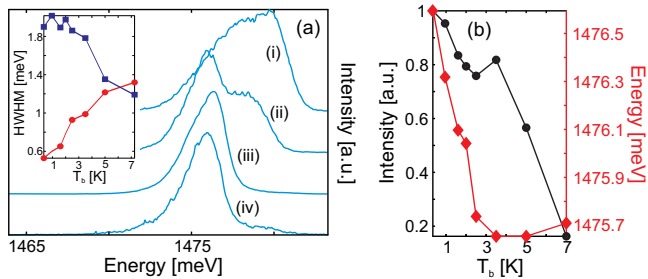}
\caption{(Color Online): (a) Photoluminescence spectrum taken: in the
illuminated region (i), 10 $\mu$m away from the laser spot (ii),
at the position of the ring (iii) and 10 $\mu$m outside the ring (iv). The
inset shows the half width at half maximum (HWHM) of the photoluminescence spectrum emitted at the position
of the ring and as a function of $T_\mathrm{b}$. The low and high energy sides are shown by filled squares and circles respectively.
(b): Energy and intensity of the photoluminescence emitted at the position of the ring as a function of $T_\mathrm{b}$, diamonds and circles respectively.}
\end{figure}

In two-dimensional systems, the formation of large diameter ring-shaped
photoluminescence patterns around a laser excitation has been
the subject of intensive research since the seminal
observations by Butov et al. \cite{Butov_02} and Snoke et al. \cite{Snoke_02}.  
Studies have then concluded that the luminescence ring marks the charge
separation between electron-rich and hole-rich regions
\cite{Butov_2003,Snoke_2003,Butov_2004,Rapaport_2004}: the hole-rich region
results from the intense photo-excitation
that
leads to a high density of holes in the laser spot.
Accordingly, 
photo-injected holes diffuse away from
the illuminated region. On the other hand, single or double quantum wells
embedded in a field-effect device confine a 2D electron gas which results from
the structure's modulation doping. These electrons then flow towards the laser
spot, i.e. towards the hole-rich region, and a sharp boundary spontaneously forms between the domains
of majority charges. The macroscopic ring forms at this
interface and
marks the recombination of indirect excitons made of well
thermalized electrons and holes
\cite{Butov_2003,Snoke_2003,Butov_2004,Rapaport_2004}. 


For our experiments, the above mechanism is qualitatively supported by the
spectrum of the photoluminescence emission. As shown in Figure 2.a, at the
position of the laser excitation (i) the photoluminescence is spectrally broad,
with a full width at half maximum (FWHM) of $\approx$ 10 meV. This indicates
that electronic carriers are mostly dissociated in this region and from the
emission's spectral width we estimate a carrier density of $\approx$
5 10$^{11}$ cm$^{-2}$. On the other hand, away from the illuminated region (ii) the
photoluminescence exhibits an additional sharp peak on the low
energy side. We attribute this spectral profile to the recombination of
both indirect
excitons and unbound electron-hole pairs, the excitonic
emission lying at a lower energy. Finally, at the position of the ring (iii), but also
in its vicinity (iv), we show in Fig.2.a that the photoluminescence spectrum reduces
to the low energy emission line. This signals that these regions
are dominantly populated by indirect excitons which emit a spectrally narrow
fluorescence ($\approx$ 2.5 meV FWHM at $T_\mathrm{b}$= 340 mK). 

The temperature dependence of the photoluminescence spectrum characterizes
further
the spontaneously formed indirect excitons. Indeed, we observed at the
position of the ring, but also in its vicinity, that the emission
spectrum developed an asymmetry initiated at T$_\mathrm{b}\leq$
4K. Precisely, the inset in
Fig.2.a shows that the high energy side of the emission
is narrowed while $T_b$ is decreased whereas
the low energy side is broadened. Such
spectral asymmetry has been reported  very
recently in microscopic traps \cite{Schinner_2011,Kotthaus_2012}.
Its establishment at very
low temperatures was then interpreted as a manifestation for strong dipolar
correlations between indirect excitons at densities of the order
of 10$^{10}$ cm$^{-2}$ \cite{Schinner_2011}. For our experiments,
the exciton
density also lies in this range in the vicinity of the ring. Indeed, the energy
of the photoluminescence is blue-shifted by $\approx$ 1 meV when
$T_\mathrm{b}$ is lowered from 7 to 0.34
K (see Fig.2.b). We attribute this energy shift to the repulsive dipolar
interactions distinctive to indirect excitons.
Following latest theoretical works \cite{Schindler_08,Rapaport_09}
we then estimate that the exciton concentration is increased by $\approx$
10$^{10}$ cm$^{-2}$ at the position of the ring between $T_\mathrm{b}$= 7 and
340 mK.


The measurements reported in Figure 1 also reveal a
striking physical property of cold indirect excitons, namely that they form
macroscopic rings that undergo a spatial fragmentation into
microscopic beads for bath temperatures
below a few Kelvin.
To the best of our knowledge, this phenomenon has solely been reported 
by Butov and co-workers in a series of
experiments performed on double quantum well heterostructures
\cite{Butov_02,Yang_2006,Yang_2010,High_2012}. Theoretical works
have then addressed this puzzling transition: approaches
relying on the Bose statistics of
dipolar excitons have concluded that
the fragmentation indicates the formation of a quantum degenerate gas
\cite{Levitov_2005,Liu_2006}. By
contrast, other theoretical works have underlined that classical processes maybe
sufficient to induce the fragmentation
\cite{Sughakov_2006,Sughakov_2007,Ivanov_2012}.

To further study the fragmentation of the exciton pattern,
we  performed  spatially resolved interferometric measurements in order to
deduce the excitons spatial coherence in the vicinity of the ring.
For that, we magnified the emission and sent it towards a 
Mach-Zehnder interferometer aligned close to the optical contact. Therefore, 
our measurements were not affected by the temporal coherence of indirect 
excitons.
The magnified photoluminescence emission was then split between the arms 1 and 2
of our
interferometer, and, as in previous works \cite{High_2012,High_2012b}, we
introduced a vertical tilt angle
($\alpha$)
between the output 
of the two arms. Hence, interference fringes are aligned horizontally and
$\alpha$ was set such that the
interference's period was equal to $\approx$ 2.5 $\mu$m. Finally, we shifted
horizontally the outputs produced by
the two
arms with respect to each other (by $\delta$x) and then measured the ouput
intensity I$_{1}$ and
I$_2$ of individual arms, and the interfering signal I$_{12}$ when both arms
are overlapped.

We used the output of our shift-interferometer, I$_{12}$, to quantify the
spatial coherence of optically active exciton states. Indeed, the
interference signal can be modelled as 
I$_{12}$(\textbf{r})=$\langle|\psi_0(\textbf{r})+e^{iq_\alpha
y}\psi_0(\textbf{r}+\delta x)|^2\rangle$ where $\psi_0$(\textbf{r}) is the
wave-function of bright excitons, $\langle..\rangle$ denotes a time
averaging, \textbf{r}=(x,y) is the coordinate in the plane of the quantum well
while
$q_\alpha$=2$\pi$sin$(\alpha)/\lambda$
with $\lambda$ equal to the emission wavelength. The 
normalized interference pattern 
I$_\mathrm{int}$=(I$_{12}$-I$_{1}$-I$_2$)/2$\sqrt{I_{1}I_2}$ then 
reveals
the first order coherence function of indirect excitons,
$g^{(1)}_\mathrm{\textbf{r}}(\delta
x)$=$\langle\psi^*_0(\textbf{r})\psi_0(\textbf{r}+\delta
x)\rangle$/$\sqrt{\langle|\psi_0(\textbf{r})|^2\rangle\langle|\psi_0(\textbf{r}
+\delta x)|^2\rangle}$,
since I$_\mathrm{int}$(\textbf{r})=cos($q_\alpha
y+\phi_\mathrm{\textbf{r}}$)$|g^{(1)}_\mathrm{\textbf{r}}(\delta x)|$ where
$\phi_\mathrm{\textbf{r}}$=arg($g^{(1)}_\mathrm{\textbf{r}}$). 
Hence, the interference's visibility is controlled by the degree 
of spatial coherence of bright excitons, while the position of the interference
fringes
reveals the phase of the $g^{(1)}$-function, i.e. the phase difference
between the interfering excitonic wave-functions.

\begin{figure}
\includegraphics[width=.9\columnwidth]{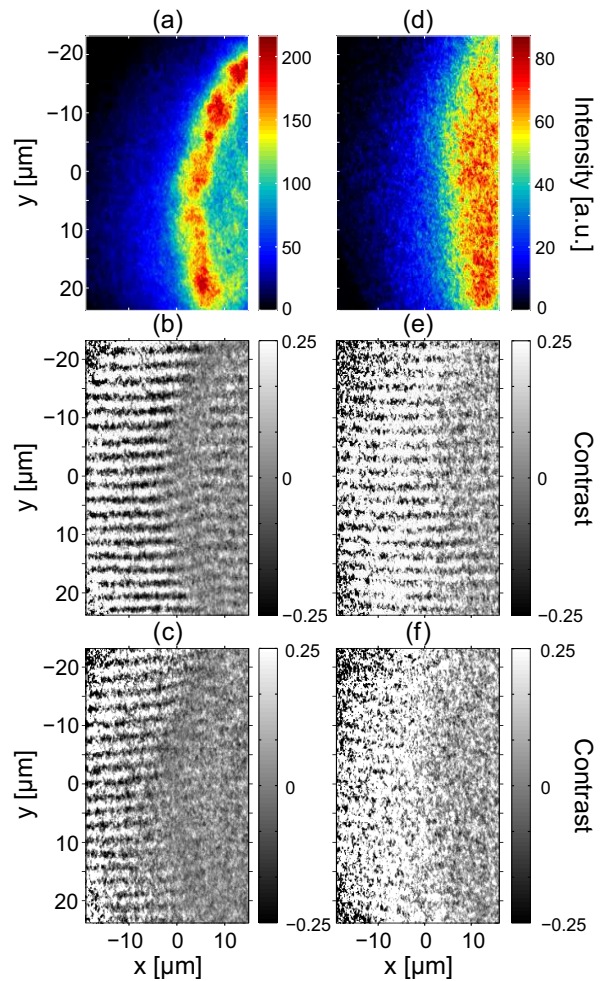}
\caption{(Color Online): The left column displays the ouput image  from the
fixed
arm of the interferometer (a) and the resulting interferograms I$_\mathrm{int}$
measured at 340 mK and for lateral shifts
$\delta$x=1.4 (b) and 2.5 $\mu$m (c). The right column (d-f) displays the
same experimental results obtained at a bath temperature T$_\mathrm{b}$=7K.}
\end{figure}

In Figure 3 we present the results of our interferometric measurements. At the
lowest bath temperature (T$_\mathrm{b}$=340 mK) and for a
lateral shift $\delta$x=1.4 $\mu$m
(Fig. 3.b), I$_\mathrm{int}$ displays clear interference fringes in the region
outside of the fragmented ring while  
at the position of the ring itself and in the inner part the signal is more
blurred. Interestingly,  we note that the interference pattern bends at the
position of the ring while it is horizontally aligned on the left and right
hand-side, as imposed by the vertical tilt in the interferometer. This signals
that the bright excitons' wave-function experiences a phase shift across the
fragmented ring, by approximately $\pi$
in these measurements. Increasing the lateral shift to 
$\delta$x=2.5 $\mu$m (Fig. 3.c) almost annihilates this structure while 
fringes persist in the outer region of the ring.
By contrast, we observed that the interference pattern does not exhibit such a
rich structure at T$_\mathrm{b}$=7K, 
i.e. when the exciton ring is no longer fragmented (see the right column in Fig.
3). In general, the interference visibility
is reduced compared to the low temperature regime, the interference pattern 
vanishing for a lateral shift $\delta$x=2.5 $\mu$m.

\begin{figure}
\includegraphics[width=\columnwidth]{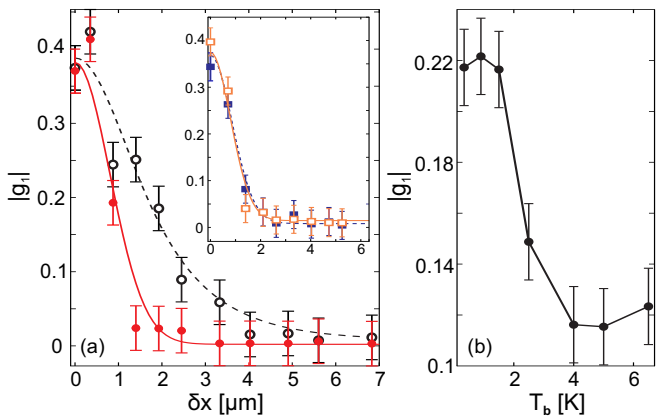}
\caption{(Color Online): (a): Inteference visibility $|g^{(1)}|$ as a function
of the lateral shift $\delta$x of the Mach-Zehnder inteferometer. Measurements
were realized at T$_\mathrm{b}$=340 mK
and at the position of the ring (filled cirles) and 10 $\mu$m outside (open
circles). The inset diplays the same measurements at T$_\mathrm{b}$= 7K, filled
and 
and open squares respectively. Solid and dash lines show the best fit of our
theoretical model to the experimental data.
(b): Variation of $|g^{(1)}|$ measured 10 $\mu$m outside the ring as a function
of T$_\mathrm{b}$ for $\delta$x=1.8 $\mu$m.}
\end{figure}

In Figure 4.a we show the variation of the interference visibility at the
position of the exciton ring but also at a
distance of 10 $\mu$m outside the latter. At T$_\mathrm{b}$=340 mK, these
data confirm quantitatively that the interference contrast  drops
more rapidly at the
position of the ring than in its outer region. There, spatial coherence
extends beyond our instrumental resolution (of $\approx$ 1.5 $\mu$m) with
$|g^{(1)}|\approx$15$\%$ for $\delta$x=2 $\mu$m. To estimate the degree
of spatial coherence, we modeled phenomenologically the variation of
$|g^{(1)}|$ by the
convolution between a Gaussian profile
with a 1.5 $\mu$m full-width at half-maximum and an exponential with a decay
constant $\xi$, the excitons coherence length. The former function accounts for
our instrumental resolution while the latter function provides the theoretical
variation of the $g^{(1)}$-function \cite{Fogler_2008,Semkat_2012} ($|g^{(1)}(\delta
x)|\propto e^{-\delta x/\xi}$). Hence, for the best fit to
our experimental data
we estimate that $\xi\approx$ 1.3 $\mu$m at a distance of 10
$\mu$m outside of the ring. By contrast, at the position of the ring we deduce a
coherence length that is limited by our spatial resolution and we
estimate that $\xi\leq$ 200 nm (see solid and dash lines in Fig. 4.a). On the
other hand, at
T$_\mathrm{b}$=7K, the interference
visibility decays rapidly at the position 
of the exciton ring but also in the outer region (see inset in Fig. 4.a):
$|g^{(1)}|$ exhibits a similar decay at these positions and we
estimate that $\xi$ does not exceed 200 nm at T$_\mathrm{b}$=7K.

Theoretical calculations
are not available yet to model the coherence length of
indirect excitons. We then
compare our estimations of $\xi$
to the case of an ideal exciton gas for which the thermal de
Broglie wavelength reads $\Lambda_\mathrm{dB}$=$h/\sqrt{2\pi
mk_BT}$, $m\approx$0.2$m_0$ being the excitons mass ($m_0$
is the electron's mass). For bath temperatures ranging from 0.34 to 7K,
$\Lambda_\mathrm{dB}$ then
varies from 160 to 40 nm. At 340 mK and 10 $\mu$m outside the fragmented ring
$\xi\gg\Lambda_\mathrm{dB}$ unambiguously signals the appearance of spatial
coherence beyond the textbook situation of non-interacting particles.  On the
other hand, at the ring's position, but also
in the higher temperature regime (7K), we observed that
$\xi\sim\Lambda_\mathrm{dB}$ with our experimental resolution which does not
allow us to draw clear conclusions.

In a last experiment, we studied the establishment of extended coherence outside
the ring while the bath temperature is lowered. Precisely, we measured the 
interference visibility as a function of 
T$_\mathrm{b}$ for a fixed lateral shift $\delta$x=1.8 $\mu$m  and 10 $\mu$m
outside the macroscopic ring. As shown in Fig.4.b, we then observed two
regimes at low and high temperatures respectively. In the latter one
(T$_\mathrm{b}\geq$4 K) the interference contrast remains constant
($|g^{(1)}|\approx$12 $\%$) while at lower
temperatures (T$_\mathrm{b}\leq$4 K) it
abruptly increases to $\approx$22 $\%$ \cite{note}. This behavior reveals that
extended coherence is solely established in the regime where the
ring of indirect excitons is fragmented, i.e. below a few Kelvin. At higher
temperatures, we attribute the non-vanishing interference visibility to our
experimental resolution \cite{Semkat_2012}, however we can not exclude that
long-range correlations persist in this regime.

To summarize, we have shown that spatially indirect excitons can realize a
macroscopically
coherent state in the regime where they form a ring shaped pattern fragmented in
microscopic beads. 
Particularly, our experiments agree with latest studies by Butov et
al. \cite{High_2012} and signal that extended coherence is established in
the vicinity of the ring fragments. Interestingly, the
photoluminescence intensity in this region is weak compared to that of
microscopic beads. This behavior may relate to the striking physical properties of indirect
excitons at very low temperatures, as underlined in recent theoretical works \cite{Rapaport_09,Roland_2012}.

This work was supported financially by the EU-ITN INDEX, by the
Spanish MEC (TOQATA), by the Spanish MINECO (Grant
TEC2011-29120-C05-04), CAM (Grant S2009ESP-1503) and by the ERC AdG QUAGATUA. F.D. also acknowledges
the Ramon y Cajal program. Furthermore M.A. and F.D. are grateful to M.
Lewenstein for his continuous support.



\begin{thebibliography}{99}
\bibitem{note} these contrasts are $\approx$
5$\%$ greater than for the measurements shown in Fig.3 which we attribute to a
better alignment of our interferometer
\end{thebibliography}
\end{document}